\documentclass[pra,graphicx,twocolumn]{revtex4-1}

\usepackage{graphicx}
\DeclareGraphicsExtensions{.pdf,.png,.jpg,.eps}

\draft 

\begin{document}


\title{Reconciliation of the Rosen and Laue theories of special relativity in a linear dielectric medium} 



\author{Michael E. Crenshaw}
\affiliation{Charles M. Bowden Research Laboratory, US Army Combat Capabilities Development Command (DEVCOM) - Aviation and Missile Center, Redstone Arsenal, AL 35898, USA}


\date{\today}

\begin{abstract}
The theory of dielectric special relativity was derived by Laue from
a fundamental physical basis in Einstein's special relativity and the
relativistic velocity sum rule.
The Laue theory is experimentally verified by the Fizeau water tube
experiment.
In contrast, the Rosen version of dielectric special relativity was
derived heuristically and has no experimental validation.
Consequently, the Rosen theory and its consequences are mostly ignored
in the scientific literature and there is little to no discussion about
the incompatibility of the two theories of relativity in a dielectric.
In this article, the Laue theory is developed from boundary conditions
using inertial reference frames moving at constant velocity along the
interface between a simple linear dielectric medium and the vacuum.
Then, the Rosen theory is derived in the context of inertial frames of
reference moving at constant velocity in the interior of an arbitrarily
large linear isotropic homogeneous dielectric medium.
These derivations show that the Laue and Rosen theories of dielectric
special relativity are both correct but have different regimes of
applicability.
The Rosen theory applies to physics in the interior of a simple linear
dielectric and the Laue theory is used to relate these physics to a
Laboratory Frame of Reference in the vacuum where measurements can
be performed.
\end{abstract}


\maketitle 

\par
\section{Introduction}
\par
In 1907, Laue \cite{BILaue,BILaue2} applied the Einstein relativistic
velocity sum rule to a transparent block of dielectric with macroscopic
refractive index $n$ that is located in the vacuum.
When the dielectric is at rest in the Laboratory Frame of Reference the
speed of light in the dielectric is $w=c/n$.
When the dielectric block is moving at velocity ${\textbf v}$ in the
Laboratory Frame of Reference, the speed of light in the moving
dielectric medium is given by the Einstein relativistic velocity sum
rule as \cite{BILaue}
\begin{equation}
w^{\prime}=\frac{
\sqrt{v^2+\frac{c^2}{n^2}+2v\frac{c}{n}\cos\theta
-\frac{v^2}{n^2}\sin^2\theta}
}{
1+\frac{v}{cn}\cos\theta
} \, ,
\label{EQv1.01}
\end{equation}
where $\theta$ is the angle between the direction of light propagation
and the direction in which the dielectric is moving.
In two simple limiting cases, we have \cite{BILaue}
$$
w^{\prime}
=\frac{\frac{c}{n}\pm v}{1\pm\frac{v}{cn}}
$$
\begin{equation}
=\frac{c}{n}+ \left ( 1-\frac{1}{n^2} \right )
\left (
\pm v-\frac{v^2}{cn}\pm \frac{v^3}{c^2n^2}- \frac{v^4}{c^3n^3}\pm\cdots
\right )
\label{EQv1.02}
\end{equation}
for light propagating in the $+/-$ direction of the velocity of the
dielectric through the vacuum and \cite{BILaue}
$$
w^{\prime}
=\sqrt{\frac{c^2}{n^2} +v^2\left ( 1-\frac{1}{n^2}\right )}
=\frac{c}{n}+\frac{1}{2}\frac{v^2}{nc}\left ( n^2-1)\right ) -
$$
\begin{equation}
\frac{1}{2}\cdot\frac{1}{4}\frac{v^4}{nc^3}(n^2-1)^2+
\frac{1}{2}\cdot\frac{1}{4}\cdot\frac{3}{6}\frac{v^6}{nc^5}(n^2-1)^3
\cdots
\label{EQv1.03}
\end{equation}
for transverse propagation.
Dispersion is treated by using the value of $n$ corresponding to the
frequency of the light field. \cite{BILaue}
The Lorentz factor in the dielectric medium
\begin{equation}
\gamma_v= \frac{1}{\sqrt{1-v^2/c^2}}
\label{EQv1.04}
\end{equation}
is the same as the vacuum Lorentz factor.
\par
In a 1952 American Journal of Physics article, Rosen \cite{BIRosen}
considered an arbitrarily large, but finite, macroscopic Maxwellian
dielectric such that an observer in the interior of the dielectric has no
access to the vacuum
{\it i}) microscopically, because the continuum limit precludes an
interstitial vacuum, and {\it ii}) macroscopically, because the boundary
of the dielectric is too far away from the interior for light to travel
within the duration of an experiment.
Because the vacuum is inaccessible, there is no way to determine the
speed or direction of the dielectric medium with respect to the
Laboratory Frame of Reference.
Consequently,  there is no way to apply the Einstein relativistic
velocity sum rule.
Then the speed of light in a Maxwellian dielectric $w$
is independent of the motion of the source or material.
The index-dependent material Lorentz factor \cite{BIRosen,BIMP,BIFinn}
\begin{equation}
\gamma_m
=\frac{1}{\sqrt{1-v^2/w^2}}
=\frac{1}{\sqrt{1-n^2v^2/c^2}},
\label{EQv1.05}
\end{equation}
that was obtained by Rosen by replacing $c$ in the vacuum theory
with
\begin{equation}
w=c/n
\label{EQv1.06}
\end{equation}
is incommensurate with the Lorentz factor, Eq.\ (\ref{EQv1.04}), of
the Laue theory.
\par
The Laue \cite{BILaue,BILaue2} theory of special relativity in a
dielectric was firmly established as a fundamental physical principle
decades before the Rosen \cite{BIRosen} article.
Further, the Laue theory has a fundamental basis in Einstein special
relativity and the relativistic velocity sum rule and it is
experimentally verified by the Fizeau \cite{BIFizeau} water tube
experiment.
In contrast, the Rosen \cite{BIRosen} dielectric special relativity
theory was derived heuristically and the material Lorentz factor,
Eq.~(\ref{EQv1.05}), is incommensurate with the Lorentz factor,
Eq.~(\ref{EQv1.04}) of Einstein's special relativity.
Moreover, there is no experimental verification of Rosen's special
relativities.
Consequently, the Rosen theory and its consequences are almost
completely ignored in the scientific literature and there is little to
no discussion about the incompatibility of the two theories of
relativity in a dielectric.
\par
Here, we argue that the Laue and Rosen treatments of special relativity
in a dielectric are both correct but have different regimes of
applicability.
These differences in applicability are obscured by the plug-and-play
manner in which the two theories were originally derived:
{\textit{i}) There is nothing in the velocity sum rule derivation of the
Laue theory to indicate any limitations on its validity.
{\textit{ii}) Likewise, the phenomenological substitution of $w=c/n$ for
$c$ presents no obvious restrictions to impose on the Rosen theory.
\par
In this article, we derive the Laue and Rosen theories in the context of
inertial frames of reference moving at constant velocities in two
different physical configurations:
\textit{i}) The Laue theory is developed from boundary conditions
for inertial frames of reference moving at constant velocity along
the interface between a simple linear dielectric medium and the vacuum
of free space.
\textit{ii}) The Rosen theory is derived in the context of inertial
frames of reference moving at constant velocity in the interior of an
arbitrarily large linear isotropic homogeneous dielectric medium.
We conclude that the Rosen theory applies to the relativistic physics in
the interior of an arbitrarily large simple linear dielectric and the
Laue theory is used to relate these physics to a Laboratory Frame of
Reference in the vacuum where other-than-optical measurements can be
performed.
A Maxwellian dielectric with a macroscopic refractive index $n$ is
continuous at all length scales so that any non-optical measuring device,
such as a clock (no matter how small), will always displace the
dielectric. (The terrestrial atmosphere is displaced by a measuring device, but the effect can often be neglected).
\par
\section{Derivation of Laue result}
\par
We consider two inertial reference frames, $S(x,y,z)$ and
$S^{\prime}(x^{\prime},y^{\prime},z^{\prime})$, in a standard
configuration \cite{BIRindler} in which $x$ and $x^{\prime}$ are
collinear, $y$ is parallel to $y^{\prime}$, $z$ stays parallel to
$z^{\prime}$, and $S^{\prime}$ translates at a constant speed in the
direction of the negative $x$-axis.
The origins of the two systems coincide at some initial time $t_0=0$.
At each point in each coordinate system, time is measured by an
idealized clock and all the clocks in each coordinate system have been
synchronized by one of the usual methods.
\par
We require that the $x$, $x^{\prime}$, $z$, and $z^{\prime}$ axes lie
on the surface of a semi-infinite dielectric, Figs.~1 and 2.
Then the upper half-space, $y>0$ and $y^{\prime}>0$, is modeled as a
linear isotropic homogeneous dielectric with index of refraction $n$ in
the rest-frame $S$ where the speed of light is $c/n$.
(The dielectric can also be finite, we only require a plane surface for
an interface.)
The lower half-space, $y<0$ and $y^{\prime}<0$, is vacuum in which the
speed of light is $c$.
\par
At time $t_d=t_d^{\prime}=t_v=t_v^{\prime}=0$, a bi-directional light
pulse is emitted from the common origin $o$ along the $\pm y$- and
$\pm y^{\prime}$-axes.
The pulse is reflected by a mirror in the vacuum at $y=-D_v$
and returns to the origin at time $\Delta t_v=2D_v/c$.
The pulse is also reflected by a mirror in the dielectric at $y=D_d$
and returns to the origin at time $\Delta t_d=2n D_d/c$.
The locations of the mirrors are adjusted so that both reflections
return to the origin at the same time such that
\begin{equation}
\Delta t_v=\Delta t_d \, ,
\label{EQv1.07}
\end{equation}
by construction.
\begin{figure}
\includegraphics[scale=0.70]{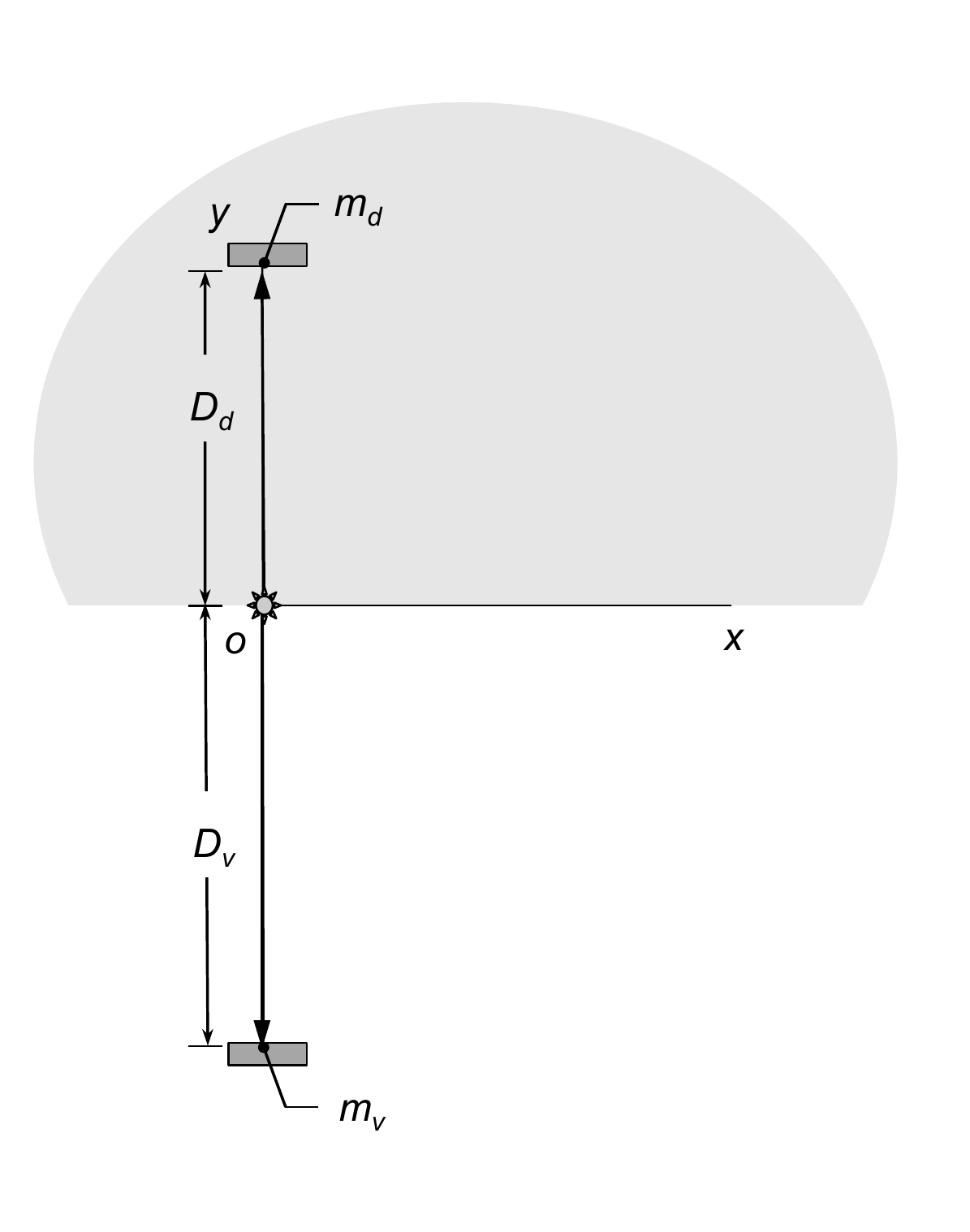}
\caption{Coordinate frame $S$ at the dielectric/vacuum boundary.}
\label{fig1}
\end{figure}
\par
The trajectory of the light pulse in the $S^{\prime}$ frame of
reference is shown in Fig.~2.
The translation of the $S^{\prime}$ frame is transverse to the $y$-axis
so the distance from the mirror at $m_v^{\prime}$ to the
$x^{\prime}$-axis is $D_v$, the same as the distance from the mirror at
$m_v$ to the $x$-axis.
Viewed from the $S^{\prime}$ frame, the light pulse is emitted from the
point $o$ at time $t_v^{\prime}=0$, is reflected from the mirror at
point $m_v^{\prime}$, and is detected at the point $d_v^{\prime}$ at
time $t_v^{\prime} =\Delta t_v^{\prime}$.
During that time, the point of emission/detection has moved a
distance $v_v\Delta t_v^{\prime}$.
Here, $v_v=dx_v^{\prime}/(dt_v^{\prime})$ is the speed of $S^{\prime}$
relative to $S$ defined in the vacuum.
\begin{figure}
\includegraphics[scale=0.70]{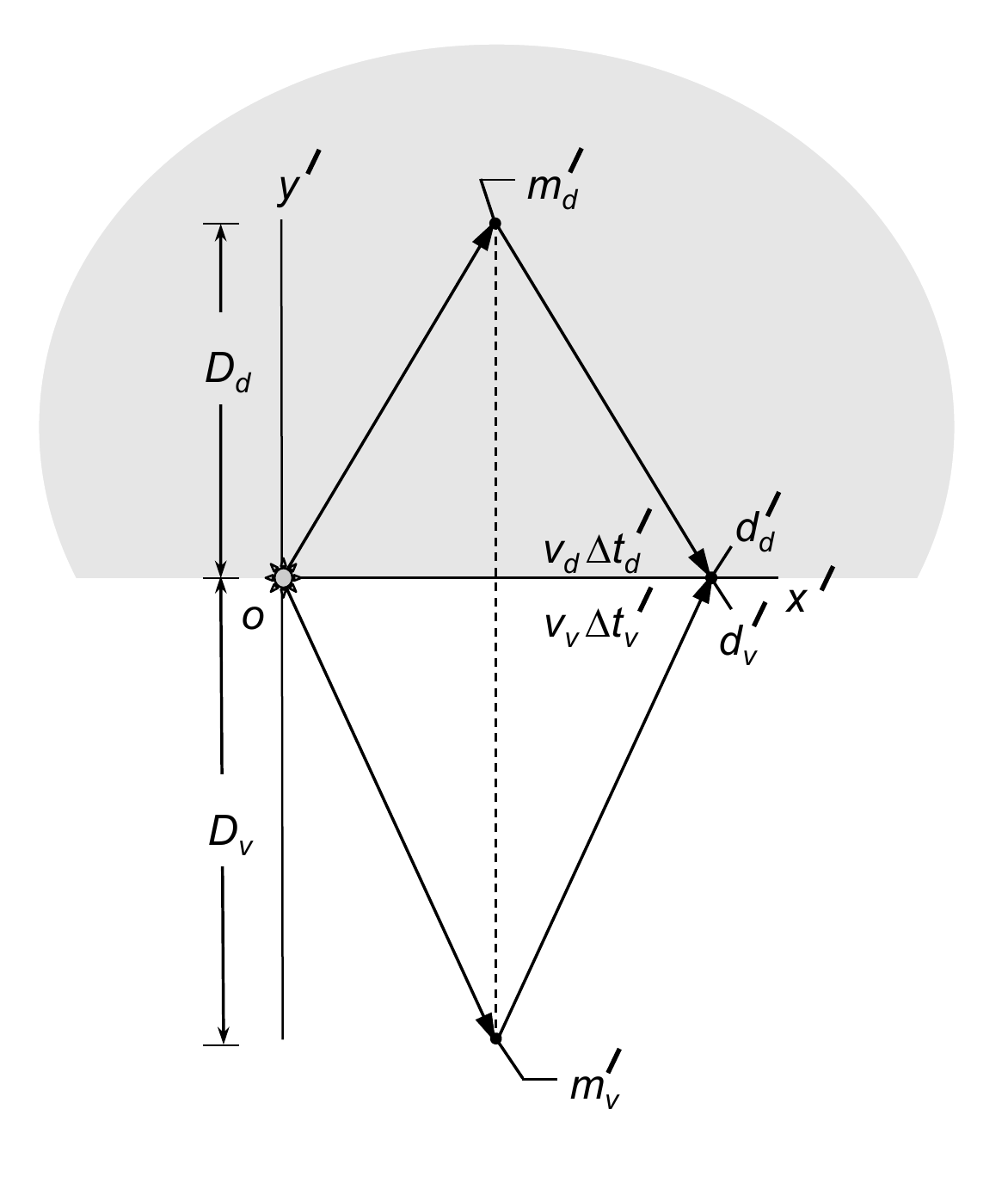}
\caption{Coordinate frame $S^{\prime}$ at the dielectric/vacuum boundary.}
\label{fig2}
\end{figure}
\par
Events that occur at both the same time \textit{and} the same place in
one inertial reference frame occur simultaneously in all inertial
reference frames.
The pulse that travels through the vacuum is reflected back to the
origin and arrives at the same time that the pulse makes a round trip
through the dielectric.
The principle of simultaneity gives us the condition
\begin{equation}
v_d\Delta t_d^{\prime} = v_v\Delta t_v^{\prime} \,
\label{EQv1.08}
\end{equation}
where $v_d=dx_d^{\prime}/(dt_d^{\prime})$ is defined by
Eq.~(\ref{EQv1.08}) to be the speed of $S^{\prime}$ relative to $S$
on the dielectric side of the interface.
\par
By symmetry, the light is reflected from the mirror at a time
$t_d^{\prime}=\Delta t_d^{\prime}/2$ making the distance the light
travels from the origin to the
mirror $c_d^{\prime}\Delta t_d^{\prime}/2$,
where $c_d^{\prime}$ is the speed of light in the direction
$\overrightarrow{om}_d^{\prime}$ in the $S^{\prime}$ frame of reference.
By the Pythagorean theorem for light in the dielectric, we have
\begin{equation}
(c_d^{\prime}\Delta t_d^{\prime})^2
=(c_d\Delta t_d)^2+(v_d\Delta t_d^{\prime})^2 \, ,
\label{EQv1.09}
\end{equation}
where we have made use of reflection symmetry about the midpoint.
We also need
\begin{equation}
\Delta t^{\prime}_v=\gamma_v\Delta t_v
\label{EQv1.10}
\end{equation}
from the vacuum theory.
\par
Regrouping terms in the Pythagorean theorem, Eq.\ (\ref{EQv1.09}), one
obtains
\begin{equation}
(\Delta t_d^{\prime})^2
 \left ( {c_d^{\prime}}^2 -v_d^2 \right )
=c_d^2\Delta t_d^2 \, .
\label{EQv1.11}
\end{equation}
Substituting Eqs.\ (\ref{EQv1.08}) and (\ref{EQv1.10}) into the
previous equation results in
\begin{equation}
\frac{v_v^2\gamma_v^2\Delta t_v^2}{v_d^2}
\left ( {c_d^{\prime}}^2 -v_d^2 \right )
=c_d^2\Delta t_d^2 \, .
\label{EQv1.12}
\end{equation}
Applying Eqs.\ (\ref{EQv1.04}) and (\ref{EQv1.07}), we obtain
\begin{equation}
{c_d^{\prime}}^2 \frac{v_v^2}{v_d^2} =
c_d^2\left (1-\frac{v_v^2}{c^2}+\frac{n^2v_v^2}{c^2}\right ) \, ,
\label{EQv1.13}
\end{equation}
where $c_d=c/n$ is the speed of light in the rest frame of the
dielectric.
Then
\begin{equation}
\frac{c_d^{\prime}\Delta t_d^{\prime}}{c\Delta t_v^{\prime}}=
\frac{1}{c}\sqrt{ \frac{c^2}{n^2}+v_v^2
\left ( 1-\frac{1}{n^2}\right )}
\label{EQv1.14}
\end{equation}
is the ratio of the distance traveled by light along the hypotenuse
in the dielectric compared to the distance in the vacuum
in Fig.~2.
We multiply Eq.\ (\ref{EQv1.14}) by $c$ and find that this analysis,
Eq.~(\ref{EQv1.15}), reproduces the Laue result, Eq.\ (\ref{EQv1.03}),
where
\begin{equation}
w^{\prime}=c_d^{\prime}\frac{\Delta t_d^{\prime}}{\Delta t_v^{\prime}}
= \sqrt{ \frac{c^2}{n^2}+v_v^2\left ( 1-\frac{1}{n^2}\right ) }
\label{EQv1.15}
\end{equation}
is the speed of light in the dielectric that is measured by an observer
in the vacuum-based Laboratory Frame of Reference.
Note that Eq.\ (\ref{EQv1.15}) displays the characteristic transverse
Fresnel drag formula, Eq.\ (\ref{EQv1.03}), that modifies the
rest-frame speed of light in a dielectric, $c/n$.
There is, however, a new interpretation in terms of a model that matches
physical quantities at the boundary between a dielectric and the vacuum
instead applying the Einstein relativistic velocity sum rule to a
dielectric block in inertial motion in a Laboratory Frame of Reference.
\par
\section{Derivation of Rosen result}
\par
In order to derive Rosen's result, we now consider a different physical
setting.
Two inertial frames of reference, $S(x,y,z)$ and
$S^{\prime}(x^{\prime},y^{\prime},z^{\prime})$, in a standard
configuration \cite{BIRindler} are located in the interior of an
arbitrarily large linear isotropic homogeneous dielectric medium,
Fig.~3.
As before, the origins of the two systems coincide at time $t_0=0$ and
all clocks are synchronized.
At time $t_d=t_d^{\prime}=0$, a light pulse is emitted from the common
origin along the positive $y$ and $y^{\prime}-$axes.
In the $S$ frame of reference, the pulse is reflected by a mirror in
the dielectric at $y=D_d$ and returns to the origin at time
$\Delta t_d=2nD_d/c$.
The trajectory of the light pulse in the $S^{\prime}$ frame of
reference is shown in Fig.~4.
The translation of the $S^{\prime}$ frame is transverse to the $y$-axis
so the distance from the mirror at $m_d^{\prime}$ to the
$x^{\prime}$-axis is $D_d$, the same as the distance from the mirror at
$m_d$ to the $x$-axis.
Viewed from the $S^{\prime}$ frame, the light pulse is emitted from the
point $o$ at time $t_d^{\prime}=0$, is reflected from the mirror at
point $m_d^{\prime}$, and is detected at the point $d_d^{\prime}$ at
time $t_d^{\prime} =\Delta t_d^{\prime}$.
During that time, the point of emission/detection has moved a
distance $v_d\Delta t_d^{\prime}$.
\begin{figure}
\includegraphics[scale=0.70]{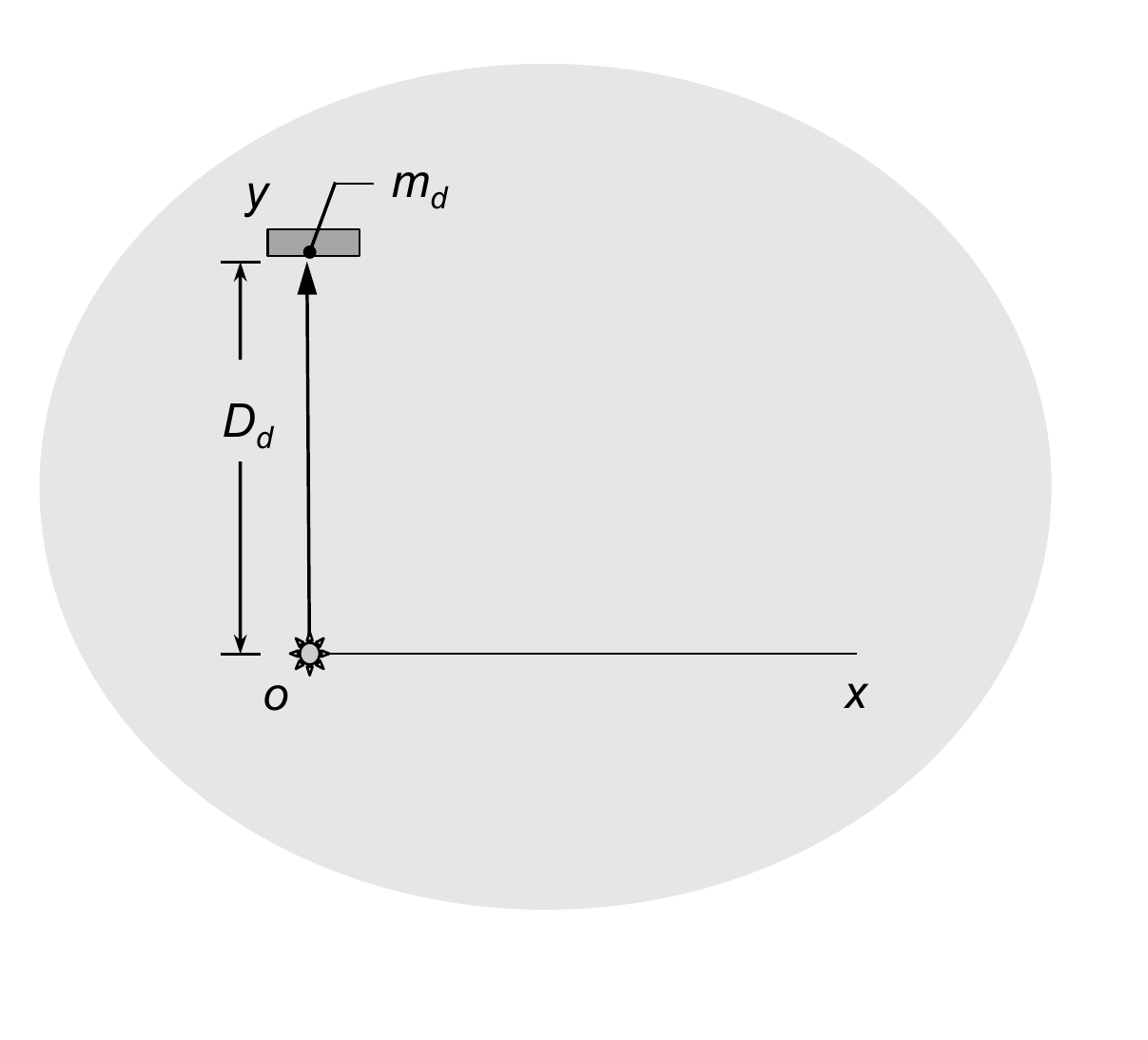}
\caption{Coordinate frame $S$ in a dielectric.}
\label{fig3}
\end{figure}
\par
By the Pythagorean theorem, we have
\begin{equation}
(c_d^{\prime}\Delta t_d^{\prime})^2
=(c_d\Delta t_d)^2+(v_d\Delta t_d^{\prime})^2 \, ,
\label{EQv1.16}
\end{equation}
where we have again used reflection symmetry about the midpoint.
We write the previous equation as
\begin{equation}
\Delta t_d^{\prime}=
\frac{\Delta t_d}{\sqrt{{c_d^{\prime}}^2/c_d^2-v_d^2/c_d^2}}
\label{EQv1.17}
\end{equation}
and define the Lorentz factor $\gamma_d$ by
\begin{equation}
\Delta t_d^{\prime}= \gamma_d \Delta t_d
\label{EQv1.18}
\end{equation}
such that
\begin{equation}
\gamma_d= \frac {1}{\sqrt{{c_d^{\prime}}^2/c_d^2-v_d^2/c_d^2}} \,.
\label{EQv1.19}
\end{equation}
\begin{figure}
\includegraphics[scale=0.70]{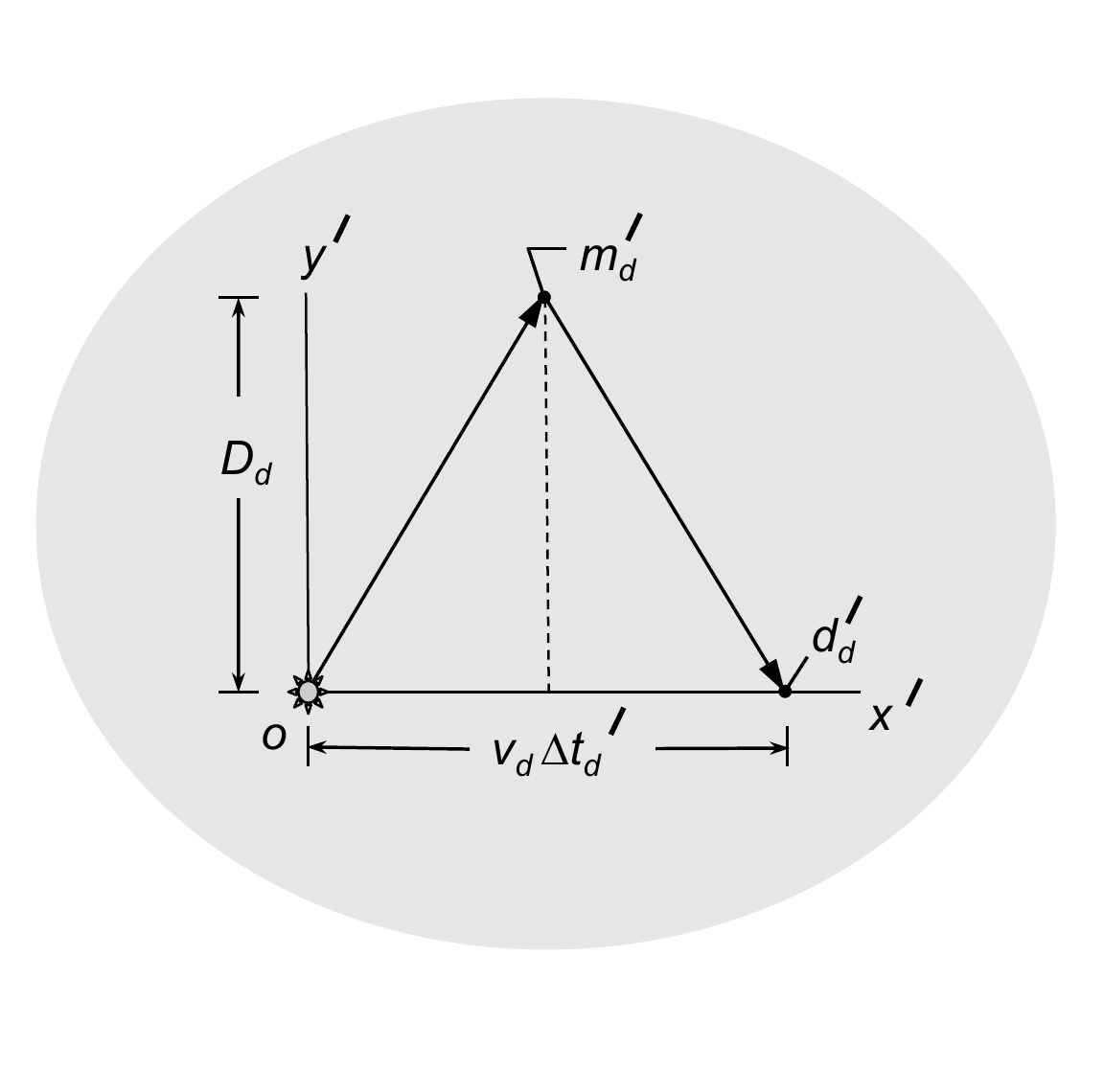}
\caption{Coordinate frame $S^{\prime}$ in the dielectric.}
\label{fig4}
\end{figure}
\par
At this point, there are more unknowns than equations and we can
proceed no further without some additional condition.
When Einstein faced the equivalent problem for free space, he postulated
that light travels at a uniform speed $c$ in the vacuum, regardless of
the motion of the source.
Here, the isotropy of an arbitrarily large homogeneous continuous
dielectric medium leads us to postulate that light travels at a uniform
speed $c_d$ in the dielectric, basically the same reasoning that led
to the Einstein postulate.
Then, we can substitute
\begin{equation}
c_d^{\prime}=c_d
\label{EQv1.20}
\end{equation}
into Eq.~(\ref{EQv1.19}) to obtain
\begin{equation}
\gamma_d= \frac {1}{\sqrt{1- v_d^2/c_d^2}} \,.
\label{EQv1.21}
\end{equation}
It can be argued that the Lorentz factor is always the vacuum Lorentz
factor, Eq.~(\ref{EQv1.03}), because the dielectric can always be
modeled as particles and interactions in the vacuum where Einstein's
special relativity is valid.
However, we are working in the continuum limit in which the material is
treated as being continuous at all length scales.
Specifically, an Einsteinian material is constructed by adding
fundamental physical entities to the vacuum, not by deconstructing a
continuous medium.
Then, in the limit of continuum electrodynamics, the macroscopic Lorentz
factor in an arbitrarily large simple
linear dielectric is given by Eq.~(\ref{EQv1.21}).
Now, the speed of light $c_d$ will be different in different
dielectrics and we are considering only simple linear dielectric
materials in which the speed of light is inversely proportional to some
constant $n$.
For each linear isotropic homogeneous dielectric with index $n_i$, there
is a different material Lorentz factor
\begin{equation}
{\gamma_m}_i= \frac {1}{\sqrt{1- n_i^2v_d^2/c^2}}
\label{EQv1.22}
\end{equation}
that corresponds to a different theory of relativity for each material
with a speed of light $c/n_i$ as suggested by Rosen. \cite{BIRosen}
Boundary conditions are used to relate dielectric special relativities
to each other and to the vacuum as the vacuum theory can be considered
to be a special case of a dielectric special relativity with $n=1$.
This was demonstrated above where we derived the Laue theory using
boundary conditions at a dielectric/vacuum interface.
\par
\section{Reconciliation of the Laue and Rosen Analyses}
\par
The Einstein theory of special relativity uses transformations
between different inertial reference frames moving at constant
velocities in vacuum. \cite{ BIRindler,BISchwarz}
Laue \cite{BILaue} showed that the Einstein relativistic velocity sum
rule explained Fizeau's \cite{BIFizeau} $19^{th}$-century experiments
of light dragging by a moving dielectric significantly contributing to
the acceptance and development of Einstein's theory of special
relativity. \cite{BIRindler,BILaue2,BIConv}
According to the Laue theory, the speed of light in a moving dielectric,
Eq.~(\ref{EQv1.01}), depends on the speed and direction of movement of
the dielectric material in the Laboratory Reference Frame.
In contradiction with this, the Rosen theory asserts that the speed of
light in a linear isotropic homogeneous dielectric is independent of the
motion of the source.
Noting that the light path through the dielectric that is shown in
Fig.~4 is the same as the light path through the dielectric that is
shown in Fig. 2, we substitute Eq.~(\ref{EQv1.20}) into
Eq.~(\ref{EQv1.15}) and use $c_d=c/n$ for a simple linear dielectric to
obtain
\begin{equation}
\frac{\Delta t_d^{\prime}}{\Delta t_v^{\prime}}
= \sqrt{ 1+\frac{v_v^2}{c^2}\left ( n^2-1 \right ) } \, .
\label{EQv1.23}
\end{equation}
Then the measurement of an increment of time is affected by the
refractive index and the motion of the dielectric in the Laboratory
Frame of Reference in which the measurement is made.
While the fundamental speed of light in the moving dielectric is always
$c_d^{\prime}=c_d=c/n$, the specifics of the measurement dictate what
velocity is observed.
The apparent velocity of the light in the medium, as measured in
the vacuum-based Laboratory Frame of reference
\begin{equation}
w^{\prime}=\frac{c}{n}\frac{\Delta t_d^{\prime}}
{\Delta t_v^{\prime}}\, ,
\label{EQv1.24}
\end{equation}
depends on the refractive index of the dielectric and the motion of
the dielectric block in the frame of reference in which the measurement
is being performed.
\par
\section{Conclusion}
\par
The significance of this article is that it rehabilitates the Rosen
theory so that it can be used in situations in which physical theory is
being applied inside a dielectric as opposed to situations in which
measurements are being made in a vacuum-based (terrestrial
atmosphere-based) Laboratory Frame of Reference.
\par

\end{document}